\documentclass[%
twocolumn,
superscriptaddress, 
showpacs,
showkeys,
amsmath, amssymb, 
aps, 
prb,
]{revtex4-2}
\usepackage[colorlinks]{hyperref}
\usepackage[colorinlistoftodos]{todonotes}
\usepackage{graphicx}
\usepackage{dcolumn}
\usepackage{bm}
\usepackage{hyperref}
\usepackage{marginnote}
\usepackage{enumitem}
\hypersetup{colorlinks=true, citecolor=blue, urlcolor=blue, linkcolor=blue}
\begin{document}
\title{A Field Theory Framework of Incompressible Fluid Dynamics}
\author{Jianfeng Wu}
\affiliation{School of Aeronautics and Astronautics, Zhejiang~University,~China}
\author{Lurong Ding}
\affiliation{School of Aeronautics and Astronautics, Zhejiang~University,~China}
\author{Hongtao Lin}
\affiliation{School of Aeronautics and Astronautics, Zhejiang~University,~China}
\author{Qi Gao} 
\email{qigao@zju.edu.cn}
\affiliation{School of Aeronautics and Astronautics, Zhejiang~University,~China}%
\date{\today}

\begin{abstract}
This study develops an effective theoretical framework that couples two vector fields: the velocity field $\mathbf{u}$ and an auxiliary vorticity field $\boldsymbol{\xi}$. Together, these fields form a larger conserved dynamical system. Within this framework, the incompressible Navier-Stokes (NS) equation and a complementary vorticity equation with negative viscosity are derived. By introducing the concept of light-cone vorticity $\boldsymbol{\eta}_\pm = \mathbf{w} \pm \boldsymbol{\xi}$, the paper constructs a unified framework for coupled dynamics. Furthermore, it explores the mechanism of spontaneous symmetry breaking from $SU(2)$ gauge theory to $U(1) \times U(1)$, which leads to the emergence of the coupled vector field theory in the non-relativistic limit. This approach uncovers a connection between fluid dynamics and fundamental gauge theories, suggesting that the NS equations describe a subsystem where dissipation results from energy transfer between the velocity and auxiliary fields. The study concludes by linking the complete dynamical framework to the Abrikosov-Nielsen-Olesen-Zumino (ANOZ) theory, a non-Abelian generalization of Bardeen-Cooper-Schrieffer (BCS) theory, offering new insights into fluid dynamics and quantum fluid theory.
\end{abstract}


\maketitle
\section*{Introduction}
Since the 19th century, fluid mechanics has endeavored to establish a theoretical framework to describe complex natural phenomena, particularly turbulence and vortex dynamics. The Navier-Stokes (NS) equations provide the foundation for modeling incompressible fluid flows. However, they face significant challenges, especially in dealing with energy dissipation and non-conservative phenomena \cite{batchelor1953theory}. Traditional conservation laws fail to fully explain the dissipative nature of these equations.

While the NS equations effectively capture macroscopic fluid behavior, they inherently lack a well-defined Lagrangian formulation due to the presence of dissipation \cite{constantin2001navier}. In physics, the Lagrangian serves not only as the basis for conservation laws but also offers insights into symmetry, stability, and dynamics \cite{griffiths2008introduction}. Hence, establishing an appropriate Lagrangian formulation for the NS equations is crucial for developing a deeper theoretical understanding of energy transfer and conservation mechanisms within fluid systems.

As von Neumann emphasized \cite{badin2017variational}, variational methods are essential for analyzing the symmetry, stability, and structure of fluid dynamics. A well-constructed Lagrangian can integrate physical intuition and experimental observations within a unified mathematical framework. This formalism provides new perspectives on complex phenomena such as vortex generation and the transition to turbulent flows. Within such a Lagrangian framework, it becomes more straightforward to explore nonlinear behaviors, critical phenomena, and phase transitions while systematically analyzing symmetry-breaking and stability issues.

However, in 1929, Millikan \cite{millikan1929steady} demonstrated a limitation: the NS equation's convective term, $\mathbf{u} \cdot \nabla \mathbf{u}$ (where $\mathbf{u}$ denotes velocity), cannot be derived from a singularity-free Lagrangian. This result, known as Millikan’s No-Go theorem, highlights the inherently non-conservative nature of the NS equations. Without viscosity, the NS equations reduce to the Euler equations, which possess a Lagrangian formulation. Yet, the Euler Lagrangian employs comoving displacements that intertwine time and space, diverging from standard Lagrangian mechanics. Similarly, in the absence of convection, the NS equations reduce to the Stokes equations, which also lack a conventional Lagrangian description.

Quantum fluids, by contrast, enjoy elegant Lagrangian descriptions. For instance, liquid helium-4 ($^4$He) at near-zero temperatures exhibits ideal properties such as zero viscosity and superfluidity, explainable through macroscopic quantum phenomena. The Gross-Pitaevskii (GP) theory \cite{Gross1961,Pitaevskii1961} captures $^4$He superfluidity by modeling it as a Bose-Einstein condensate (BEC) \cite{bose1924plancks,einstein1925quantentheorie} formed due to symmetry breaking. However, the GP theory cannot describe fermionic systems like helium-3 ($^3$He), whose superfluidity requires the Bardeen-Cooper-Schrieffer (BCS) theory \cite{BCS1957,Leggett2006}, wherein fermions form Cooper pairs. These pairs act as bosonic quasiparticles, undergoing condensation into a macroscopic state.

Interestingly, quantum field theory provides a general framework for handling nonlinearity. Although the Gross-Pitaevskii (GP) equation is nonlinear, auxiliary fields introduced via Hubbard-Stratonovich transformations \cite{stratonovich1957method,hubbard1959calculation} can linearize the system. Conversely, the BCS theory, although based on linear fermionic equations, leads to the nonlinear Ginzburg-Landau effective theory \cite{ginzburg1950theory} for superconductivity. This reflects a powerful principle: nonlinear systems can be reformulated as linear ones with additional degrees of freedom, and reducing those degrees restores nonlinearity in the effective theory.

Quantum fluid theories further suggest potential connections to classical fluid mechanics. In general, quantum theories can yield classical counterparts through the Wentzel-Kramers-Brillouin (WKB) approximation \cite{wentzel1926,kramers1926,brillouin1926}. Another approach involves Madelung’s transformation \cite{Madelung1926}, which relates fluid dynamics to quantum mechanics. Recent developments, such as the hydrodynamic Schrödinger equation (HSE) proposed by Meng and Yang \cite{MY2023}, have utilized this idea to represent classical hydrodynamics using quantum state functions, thereby linking classical fluid mechanics to the Ginzburg-Landau framework \cite{meng2024simulatingunsteadyfluidflows}.

These insights propose the possibility that the NS equations describe only a subset of a larger, conservative physical system. Inspired by gauge theories, one can envision a non-Abelian framework in which NS dynamics emerge from the symmetry breaking of a unified system. In this framework, two coupled fields—one representing velocity and the other adjoint vorticity—capture the complete dynamics. The incompressible NS equations arise as a non-relativistic limit of this broader framework, with scalar fields driving symmetry breaking and governing fluid behavior. This unified perspective suggests that the apparent non-conservative nature of the NS equations may reflect an incomplete understanding, and a more comprehensive theoretical framework could uncover hidden symmetries and conservation laws underlying fluid mechanics.

This study begins by exploring an effective theory coupling two vector fields: the velocity field $\mathbf{u}$ and an \textbf{adjoint vorticity field} $\boldsymbol{\xi}$. Together, these two fields form a larger dynamical system with conserved properties. The corresponding Lagrangian is simple in form, facilitating the derivation of both the NS equations for incompressible fluids and an adjoint vorticity equation exhibiting \textbf{negative viscosity}. By coupling the vorticity field $\mathbf{w} = \nabla \times \mathbf{u}$ with the adjoint vorticity $\boldsymbol{\xi}$, we introduce the concept of \textbf{light-cone vorticity}, thereby establishing a comprehensive framework for the coupled dynamics.

Once the effective theory for the two coupled vector fields is constructed, a natural question arises: \textit{Can a symmetry-breaking mechanism yield the coupled theory from a more fundamental gauge theory?} This study demonstrates that starting with an $SU(2)$ gauge theory, dynamic polarization leads to symmetry breaking into a $U(1) \times U(1)$ theory. By taking the non-relativistic limit of the resulting fields, the NS equations and the adjoint vorticity equation are naturally recovered.

\subsection{Physics Background and New Perspectives}

\subsubsection{Non-Conservativity and Subsystem Reduction in Fluid Dynamics}

The classical NS equations are central to fluid mechanics, describing key phenomena such as energy dissipation, vortex generation, and turbulence. However, a significant limitation of the NS equations is their \textbf{non-conservativity}: energy dissipates over time due to viscosity, which seems irreconcilable with classical mechanics.

We propose that the non-conservativity of the NS equations arises from the \textbf{reduction of a larger dynamical system}. The NS equations describe only a \textbf{subsystem} extracted from a more extensive system with higher conservation laws. While energy may not be conserved within the NS framework, it is transferred between the velocity field $\mathbf{u}$ and the adjoint vorticity field $\boldsymbol{\xi}$, ensuring conservation within the larger system.

\subsubsection{Adjoint Vorticity Field: Extending the Conservation Framework}

The \textbf{adjoint vorticity field} $\boldsymbol{\xi}$ is introduced to account for the energy dissipation observed in the NS equations. This field represents the flow of energy beyond what is captured by the velocity field alone.

\begin{itemize}
    \item \textbf{Physical Interpretation}: The velocity field $\mathbf{u}$ and the adjoint vorticity $\boldsymbol{\xi}$ form a complete system where energy exchange occurs between these fields. While the subsystem governed by the NS equations shows energy dissipation, this dissipation corresponds to energy transfer to the adjoint field $\boldsymbol{\xi}$, maintaining global energy conservation.
    \item\textbf{Role of the Adjoint Vorticity Field}: The adjoint field acts as a \textbf{negative viscosity field}, compensating for energy loss in the primary system. It suppresses the spread of vorticity and supports complex energy flows throughout the system. Although the adjoint vorticity field is introduced to maintain the conservation properties of the fluid system, it carries profound physical meaning. Just as the introduction of phonon dynamics \cite{frohlich1954} complements electron dynamics in the theory of conductors, revealing the microscopic mechanism of electrical resistance, the physical role of the adjoint vorticity field parallels that of phonons in conductors.
\end{itemize}

\subsubsection{Light-Cone Vorticity: A Complete Description of Coupled Dynamics}

To further describe the interaction between the velocity and adjoint vorticity fields, we introduce the concept of \textbf{light-cone vorticity}:

\begin{align}
\boldsymbol{\eta}_{\pm} = \mathbf{w} \pm \boldsymbol{\xi}.
\end{align}

This formulation provides a complete set of equations governing the coupled dynamics.

\begin{itemize}
    \item \textbf{Physical Significance of Light-Cone Vorticity}: The fields $\boldsymbol{\eta}_{+}$ and $\boldsymbol{\eta}_{-}$ represent two distinct modes of the coupled system. Their interactions encapsulate energy transfer dynamics, including diffusion, convection, and vortex interactions.
    \item \textbf{Key Features of the Light-Cone Vorticity Equations}: Cross-interaction terms, such as $\boldsymbol{\eta}_{-} \times \boldsymbol{\eta}_{+}$, reveal \textbf{nonlinear feedback mechanisms} between the vorticity and adjoint fields. These interactions form the foundation for understanding turbulence, vortex formation, and other complex fluid phenomena.
\end{itemize}

\subsubsection{Symmetry Breaking in Gauge Theory}

The study begins with an $SU(2)$ gauge theory, as known as the Abrikosov-Nielsen-Olesen-Zumino (ANOZ) theory \cite{Abrikosov1957, NO1973,Zumino1979}, and demonstrates how \textbf{dynamic polarization} leads to a $U(1) \times U(1)$ theory. One $U(1)$ field corresponds to the velocity field, and the other to the adjoint vorticity field. Taking the non-relativistic limit of these fields, we recover the NS equations and the adjoint vorticity equation.

\begin{itemize}
    \item \textbf{Gauge Symmetry Breaking}: This process shows that the \textbf{NS equations} and the adjoint vorticity equation naturally emerge from the symmetry-breaking of an $SU(2)$ gauge theory.
    \item \textbf{Coupling of the Two Vector Fields}: The interaction between the velocity and adjoint vorticity fields reveals deep connections between higher-order field theory and classical fluid mechanics.
\end{itemize}

\subsubsection{Complete Dynamics: ANOZ Theory}

Incorporating scalar fields into the framework leads to the \textbf{ANOZ theory}, a non-Abelian extension of the \textbf{BCS theory for ${^3\text{He}}$ superfluidity}. In the \textbf{decoupling limit}, the scalar field theory reduces to the \textbf{Schrödinger-Pauli equation}, which corresponds to the \textbf{Clebsch potential flow} \cite{yang2021clebsch,tao2021knotted} in fluid mechanics.

\begin{itemize}
    \item \textbf{Clebsch Mapping and Quantum Analogy}: The scalar field dynamics resemble the \textbf{Gross-Pitaevskii (GP) equation} in superfluidity, suggesting a potential \textbf{quantization of fluid systems}. The Schrödinger-Pauli equation (SPE) serves as the foundational equation for quantum fluid mechanics. In the paper \cite{MY2023}, based on the Clebsch mapping and the conservation equation of fluid flow, the corresponding quantum SPE equations of the Euler equations can be derived. In this article, we believe that the Clebsch mapping is likely formally equivalent to the decoupling condition of vector-scalar fields.
\end{itemize}

In the theoretical framework proposed in this paper, the relationships among the various theories can be illustrated by the following FIG.\ref{fig:framework}.

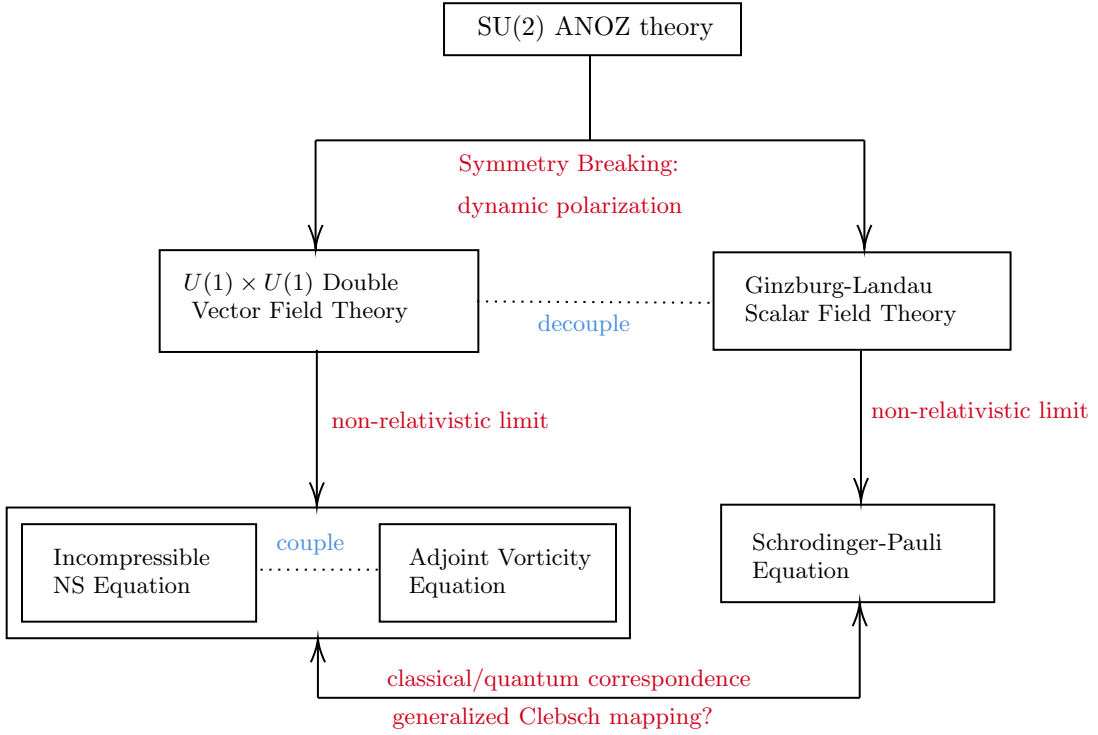
\begin{figure*}[htbp]
\centering
\tikzset{every picture/.style={line width=0.75pt}} 

\begin{tikzpicture}[x=0.82pt,y=0.82pt,yscale=-1,xscale=1]

\draw    (325,94) -- (325,133.31) ;
\draw    (198.42,133.31) -- (451.58,133.31) ;
\draw    (198.42,133.31) -- (198.42,181.31) ;
\draw [shift={(198.42,183.31)}, rotate = 270] [color={rgb, 255:red, 0; green, 0; blue, 0 }  ][line width=0.75]    (10.93,-3.29) .. controls (6.95,-1.4) and (3.31,-0.3) .. (0,0) .. controls (3.31,0.3) and (6.95,1.4) .. (10.93,3.29)   ;
\draw    (451.58,133.31) -- (451.58,181.31) ;
\draw [shift={(451.58,183.31)}, rotate = 270] [color={rgb, 255:red, 0; green, 0; blue, 0 }  ][line width=0.75]    (10.93,-3.29) .. controls (6.95,-1.4) and (3.31,-0.3) .. (0,0) .. controls (3.31,0.3) and (6.95,1.4) .. (10.93,3.29)   ;
\draw  [dash pattern={on 0.84pt off 2.51pt}]  (272.92,207.65) -- (383.17,208.31) ;
\draw    (199,230) -- (199,300.15) ;
\draw [shift={(199,302.15)}, rotate = 270] [color={rgb, 255:red, 0; green, 0; blue, 0 }  ][line width=0.75]    (10.93,-3.29) .. controls (6.95,-1.4) and (3.31,-0.3) .. (0,0) .. controls (3.31,0.3) and (6.95,1.4) .. (10.93,3.29)   ;
\draw  [dash pattern={on 0.84pt off 2.51pt}]  (173,331.5) -- (228.17,331.5) ;
\draw   (55.92,302.81) -- (343.42,302.81) -- (343.42,363.15) -- (55.92,363.15) -- cycle ;

\draw    (450,230.5) -- (450,298.15) ;
\draw [shift={(450,300.15)}, rotate = 270] [color={rgb, 255:red, 0; green, 0; blue, 0 }  ][line width=0.75]    (10.93,-3.29) .. controls (6.95,-1.4) and (3.31,-0.3) .. (0,0) .. controls (3.31,0.3) and (6.95,1.4) .. (10.93,3.29)   ;
\draw    (199.5,365.31) -- (199.5,390.65) ;
\draw [shift={(199.5,363.31)}, rotate = 90] [color={rgb, 255:red, 0; green, 0; blue, 0 }  ][line width=0.75]    (10.93,-3.29) .. controls (6.95,-1.4) and (3.31,-0.3) .. (0,0) .. controls (3.31,0.3) and (6.95,1.4) .. (10.93,3.29)   ;
\draw    (199.5,390.65) -- (449.42,390.65) ;
\draw    (449.42,390.65) -- (449.42,348.65) ;
\draw [shift={(449.42,346.65)}, rotate = 90] [color={rgb, 255:red, 0; green, 0; blue, 0 }  ][line width=0.75]    (10.93,-3.29) .. controls (6.95,-1.4) and (3.31,-0.3) .. (0,0) .. controls (3.31,0.3) and (6.95,1.4) .. (10.93,3.29)   ;

\draw    (257.67,70) -- (394.67,70) -- (394.67,94) -- (257.67,94) -- cycle  ;
\draw (271.67,74) node [anchor=north west][inner sep=0.75pt]  [font=\normalsize] [align=left] {SU(2) ANOZ theory};
\draw    (126.5,184) -- (273.5,184) -- (273.5,231) -- (126.5,231) -- cycle  ;
\draw (132.5,192.2) node [anchor=north west][inner sep=0.75pt]    {$ \begin{array}{l}
U( 1) \times U( 1) \ \text{Double} \ \\
\ \text{Vector Field Theory}
\end{array}$};
\draw    (382,185) -- (519,185) -- (519,230) -- (382,230) -- cycle  ;
\draw (395,195) node [anchor=north west][inner sep=0.75pt]   [align=left] {Ginzburg-Landau\\Scalar Field Theory};
\draw (262.67,158) node [anchor=north west][inner sep=0.75pt]  [font=\small,color={rgb, 255:red, 208; green, 2; blue, 27 }  ,opacity=1 ] [align=left] {dynamic polarization};
\draw (299.5,212.5) node [anchor=north west][inner sep=0.75pt]  [font=\small,color={rgb, 255:red, 74; green, 144; blue, 226 }  ,opacity=1 ] [align=left] {decouple};
\draw (204.33,257) node [anchor=north west][inner sep=0.75pt]  [font=\small,color={rgb, 255:red, 208; green, 2; blue, 27 }  ,opacity=1 ] [align=left] {non-relativistic limit};
\draw    (385.5,301.5) -- (511.5,301.5) -- (511.5,346.5) -- (385.5,346.5) -- cycle  ;
\draw (398.5,313.5) node [anchor=north west][inner sep=0.75pt]   [align=left] {Schrodinger-Pauli\\Equation};
\draw (453.33,252.67) node [anchor=north west][inner sep=0.75pt]  [font=\small,color={rgb, 255:red, 208; green, 2; blue, 27 }  ,opacity=1 ] [align=left] {non-relativistic limit};
\draw (263.17,138.31) node [anchor=north west][inner sep=0.75pt]  [font=\small,color={rgb, 255:red, 208; green, 2; blue, 27 }  ,opacity=1 ] [align=left] {Symmetry Breaking:};
\draw    (63,310.5) -- (171,310.5) -- (171,355.5) -- (63,355.5) -- cycle  ;
\draw (76,320.5) node [anchor=north west][inner sep=0.75pt]   [align=left] {Incompressible\\NS Equation};
\draw    (228,310.5) -- (337,310.5) -- (337,355.5) -- (228,355.5) -- cycle  ;
\draw (240,320.5) node [anchor=north west][inner sep=0.75pt]   [align=left] {Adjoint Vorticity\\Equation};
\draw (179,313.5) node [anchor=north west][inner sep=0.75pt]  [font=\small,color={rgb, 255:red, 74; green, 144; blue, 226 }  ,opacity=1 ] [align=left] {couple};
\draw (230,375.81) node [anchor=north west][inner sep=0.75pt]  [font=\small,color={rgb, 255:red, 208; green, 2; blue, 27 }  ,opacity=1 ] [align=left] {classical/quantum correspondence};
\draw (232.5,393.81) node [anchor=north west][inner sep=0.75pt]  [font=\small,color={rgb, 255:red, 208; green, 2; blue, 27 }  ,opacity=1 ] [align=left] {generalized Clebsch mapping?};

\end{tikzpicture}
\caption{The field theory framework of incompressible fluid dynamics}\label{fig:framework}
\end{figure*}

It is noteworthy that Sanders et al. recently developed a canonical Hamiltonian formulation for incompressible fluid dynamics \cite{sanders2024}. In their work, they also constructed a Lagrangian; however, their approach results in a higher-order Hamiltonian dynamical system. In its formulation, the NS equation emerges as a trivial solution to the higher-order Hamilton-Jacobi equation.

\section{The effective Lagrangian of incompressible fluid}
\paragraph*{Notational Conventions:} Quantities with a tilde denote relativistic fields or field strengths, while those without a tilde denote non-relativistic ones. Greek indices $\{\mu, \nu, \lambda\}$ refer to the Minkowski spacetime coordinates: $[0,1,2,3]$, where the $0$-direction represents time. We use Latin indices $\{a, b, c\}$ for Lie algebra space and $\{i, j, k\}$ for three-dimensional spatial coordinates. In this article, we adopt the Einstein summation convention throughout.

We denote the velocity vector field by ${\bf u}$, with its spatial components represented as $u_i, i = 1, 2, 3$. The field strength components of the velocity field ${\bf u}$ are defined as:
\begin{equation}
K_{ij} = \partial_i u_j - \partial_j u_i\,,
\end{equation}
where an additional vector field $\boldsymbol{\xi}$ is introduced along with its field strength:
\begin{equation}
f_{ij} = \partial_i \xi_j - \partial_j \xi_i\,.
\end{equation}
Now, consider the following effective Lagrangian:
\begin{equation}
\mathcal{L}_{\text{eff}} = -\xi_i \partial_t u^i + \xi_i u_j K^{ij} - \frac{\nu}{2} f_{ij} K^{ij}, \label{eq:efflag}
\end{equation}
and compute the variations with respect to the fields:
\begin{align}
\frac{\delta \mathcal{L}_{\text{eff}}}{\delta \xi_i} &= -\partial_t u^i + u_j K^{ij}, \label{eq:u1} \\
\partial_k \frac{\delta \mathcal{L}_{\text{eff}}}{\delta (\partial_k \xi_i)} &= -\nu \partial_k \partial^k u^i, \label{eq:u2} \\
\frac{\delta \mathcal{L}_{\text{eff}}}{\delta u_i} &= \xi_k K^{ki}, \label{eq:w1} \\
\partial_k \frac{\delta \mathcal{L}_{\text{eff}}}{\delta (\partial_k u_i)} &= -\nu \partial_k \partial^k \xi^i - u^k \partial_k \xi^i + \xi^k \partial_k u^i, \label{eq:w2} \\
\partial_t \frac{\delta \mathcal{L}_{\text{eff}}}{\delta (\partial_t u_i)} &= -\partial_t \xi^i. \label{eq:w3}
\end{align}
In the above variations, we neglect the contributions from $\partial_k u^k$ and $\partial_k \xi^k$. The former is omitted because we are considering incompressible fluids, and the latter follows from the manually imposed gauge-fixing condition $\partial_k \xi^k = 0$. From Eqs.~\eqref{eq:u1} and \eqref{eq:u2}, the equation of motion for the velocity field ${\bf u}$ is derived as:
\begin{equation}
\partial_t u_i = -u^j K_{ji} + \nu \partial_k \partial^k u_i, \label{eq:compovelocity}
\end{equation}
which can be rewritten in vector form as:
\begin{equation}
\frac{\partial {\bf u}}{\partial t} = -{\bf u} \times (\nabla \times {\bf u}) + \nu \nabla^2 {\bf u}. \label{eq:vectvelocity}
\end{equation}
The only difference between this equation and the standard NS equation without external forces is a gradient of kinetic energy, ${\nabla(\frac{1}{2} {\bf u} \cdot {\bf u})}$, which can be absorbed by redefining the pressure term in the NS equation. Notably, the first term on the right-hand side of Eq.~\eqref{eq:vectvelocity} corresponds to the term identified in Millikan’s No-Go theorem, which states that it cannot be derived from a variational principle. However, by introducing the vector field $\xi$, Millikan’s No-Go theorem no longer poses an obstacle to deriving the NS equation from a variational principle. 

The Euler-Lagrange equation for the $\xi$-field follows from Eqs.~\eqref{eq:w1}, \eqref{eq:w2}, and \eqref{eq:w3}:
\begin{align}
-\partial_t \xi_i &= \xi^j K_{ji} + \nu \partial_k \partial^k \xi_i + u^k \partial_k \xi^i - \xi^k \partial_k u^i \nonumber \\
&= \nu \partial_k \partial^k \xi_i + u^k \partial_k \xi_i - \xi^k \partial_i u_k. \label{eq:compvorticiy}
\end{align}
In vector form, this equation reads:
\begin{equation}
-\frac{\partial \boldsymbol{\xi}}{\partial t} = \nu \nabla^2 \boldsymbol{\xi} + ({\bf u} \cdot \nabla) \boldsymbol{\xi} - (\boldsymbol{\xi} \cdot \nabla) {\bf u} + \boldsymbol{\xi} \times (\nabla \times {\bf u}). \label{eq:vectvorticity}
\end{equation}
The equation of motion for $\boldsymbol{\xi}$, Eq.~\eqref{eq:vectvorticity}, can be interpreted as a vorticity equation with negative viscosity $(-\nu)$. We refer to this as the adjoint vorticity equation (AVE) of the NS equation. 

We define the vorticity ${\bf w}$ of the velocity field ${\bf u}$ as ${\bf w} \equiv \nabla \times {\bf u}$ and introduce the ``light-cone vorticities":
\begin{equation}
\boldsymbol{\eta}_{\pm} \equiv {\bf w} \pm \boldsymbol{\xi}.
\end{equation}
The equations of motion \eqref{eq:vectvelocity} and \eqref{eq:vectvorticity} can then be expressed as the following ``light-cone" system:
\begin{eqnarray}
\frac{\partial \boldsymbol{\eta}_{\pm}}{\partial t} &=& \nu \nabla^2 \boldsymbol{\eta}_{\mp} - ({\bf u} \cdot \nabla) \boldsymbol{\eta}_{\pm}\nonumber\\ &&+ (\boldsymbol{\eta}_{\pm} \cdot \nabla) {\bf u} \pm \frac{1}{2} \boldsymbol{\eta}_- \times \boldsymbol{\eta}_+. 
\end{eqnarray}
From the above derivation, it follows that by coupling the fluid dynamics with a ``negative-viscosity" vorticity field, the dynamics can indeed be described by a Lagrangian formulation. This indicates that the incompressible fluid system, when treated alone, is not conservative. However, when coupled with the negative-viscosity vorticity, the combined system becomes conservative. Therefore, the dynamics of incompressible fluids can be viewed as a sub-dynamics of a larger conservative system, whose Lagrangian is given by Eq.~\eqref{eq:efflag}.

\section{Double Gauge Theory with Symmetry Breaking}

Although the effective Lagrangian \eqref{eq:efflag} successfully captures the dynamics of incompressible fluid flows as a sub-dynamics of a larger coupled conservative system, the process of symmetry breaking from a higher-energy theory into such an effective Lagrangian remains unclear. In this section, we demonstrate that the dynamical system of a non-Abelian $SU(2)$ gauge field theory can break via polarization into two coupled $U(1)$ gauge field theories, corresponding respectively to the velocity field $\mathbf{u}$ and the associated vorticity field $\mathbf{\xi}$.

We first introduce a relativistic non-Abelian gauge field theory:
\begin{equation}
\tilde{\mathcal{L}} = -\frac{1}{4} \tilde{\mathcal{F}}_{\mu\nu} \tilde{\mathcal{F}}^{\mu\nu}, \label{eq:so3xso3lag}
\end{equation}
where $\tilde{\mathcal{F}}_{\mu\nu} = [\tilde{\mathcal{D}}_\mu, \tilde{\mathcal{D}}_\nu]$ is the field strength, and the covariant derivative $\tilde{\mathcal{D}}_\mu$ is defined as:
\begin{equation}
\tilde{\mathcal{D}}_\mu = \partial_\mu - ie \tilde{u}_\mu^a \tau^a - ig \tilde{\xi}_\mu^b \tau^b, \label{eq:covderivative}
\end{equation}
where $\tau^a$ and $\tau^b$ are the generators of the $su(2)$ Lie algebra, satisfying $[\tau^a, \tau^b] = i \epsilon^{abc} \tau^c$.

Although \eqref{eq:covderivative} appears to represent an $SU(2) \times SU(2)$ covariant derivative, we can redefine the vector fields as:
\begin{align}
\tilde{U}_\mu = \left( \tilde{u}_\mu^a + \frac{e}{g} \tilde{\xi}_\mu^a \right) \tau^a,
\end{align}
such that \eqref{eq:covderivative} behaves like an $SU(2)$ covariant derivative. The separation into two vector fields serves to facilitate the analysis of symmetry breaking later. Explicitly, $\tilde{\mathcal{F}}_{\mu\nu}$ is:
\begin{align}
\tilde{\mathcal{F}}_{\mu\nu} &= [\tilde{\mathcal{D}}_\mu, \tilde{\mathcal{D}}_\nu] = -ie \tilde{K}_{\mu\nu} - ig \tilde{f}_{\mu\nu} \nonumber \\
&\quad - e^2 [\tilde{u}_\mu, \tilde{u}_\nu] - g^2 [\tilde{\xi}_\mu, \tilde{\xi}_\nu] - eg [\tilde{u}_\mu, \tilde{\xi}_\nu].
\end{align}

We now consider symmetry breaking from $SU(2) \rightarrow U(1) \times U(1)$. In this case, the commutators $[\tilde{u}_\mu, \tilde{u}_\nu] = [\tilde{\xi}_\mu, \tilde{\xi}_\nu] = 0$, so:
\begin{align}
\tilde{\mathcal{F}}_{\mu\nu} = -ie \tilde{K}_{\mu\nu} - ig \tilde{f}_{\mu\nu} - eg [\tilde{u}_\mu, \tilde{\xi}_\nu].
\end{align}
We obtain:
\begin{align}
\tilde{\mathcal{F}}_{\mu\nu} \tilde{\mathcal{F}}^{\mu\nu} &= -e^2 \tilde{K}_{\mu\nu} \tilde{K}^{\mu\nu} - g^2 \tilde{f}_{\mu\nu} \tilde{f}^{\mu\nu} - 2eg \tilde{f}_{\mu\nu} \tilde{K}^{\mu\nu} \nonumber \\
&\quad - 2e^2g \left[ \tilde{u}_\mu \times \tilde{\xi}_\nu \right] \tilde{K}^{\mu\nu} - 2eg^2 \left[ \tilde{u}_\mu \times \tilde{\xi}_\nu \right] \tilde{f}^{\mu\nu} \nonumber \\
&\quad + e^2g^2 [\tilde{u}_\mu \times \tilde{\xi}_\nu] [\tilde{u}^\mu \times \tilde{\xi}^\nu].
\end{align}

Next, we assume a symmetry-breaking scenario where the $U(1)$ Lie algebra generator of the $\tilde{u}$ field is:
\begin{align}
\kappa = \alpha \tau^1 + \beta \tau^2 + \gamma \tau^3, \quad \kappa^2 = (\alpha^2 + \beta^2 + \gamma^2) \mathbb{I} = \mathbb{I}.
\end{align}
For the $\tilde{\xi}$ field, we assume polarization along the $\tau^3$ direction. Under this symmetry breaking, the term $\left[ \tilde{u}_\mu \times \tilde{\xi}_\nu \right] \tilde{f}^{\mu\nu}$ vanishes, and the Lagrangian becomes:
\begin{align}
\tilde{\mathcal{F}}_{\mu\nu} \tilde{\mathcal{F}}^{\mu\nu} &= -e^2 \tilde{K}_{\mu\nu} \tilde{K}^{\mu\nu} - g^2 \tilde{f}_{\mu\nu} \tilde{f}^{\mu\nu} - 2eg\gamma \tilde{f}_{\mu\nu} \tilde{K}^{\mu\nu} \nonumber \\
&\quad - 4e^2g (\beta^2 - \alpha^2) \tilde{u}_\mu \tilde{\xi}_\nu \tilde{K}^{\mu\nu} \nonumber \\
&\quad + 2e^2g^2 \left[ \tilde{u}_\mu^2 \tilde{\xi}_\nu^2 + (\beta - \alpha)^2 (\tilde{u}_\mu \tilde{\xi}^\mu)^2 \right].
\end{align}

The unbroken part of the Lagrangian is:
\begin{equation}
\tilde{\mathcal{L}}_{\text{unbroken}} = \frac{e^2}{4} \tilde{K}_{\mu\nu} \tilde{K}^{\mu\nu} + \frac{g^2}{4} \tilde{f}_{\mu\nu} \tilde{f}^{\mu\nu} - \frac{e^2g^2}{2} \tilde{u}_\mu^2 \tilde{\xi}_\nu^2.
\end{equation}
The broken part is:
\begin{eqnarray}
\tilde{\mathcal{L}}_{\text{broken}} = \frac{eg}{2} \gamma \tilde{f}_{\mu\nu} \tilde{K}^{\mu\nu} + e^2g (\beta^2 - \alpha^2) \tilde{u}_\mu \tilde{\xi}_\nu \tilde{K}^{\mu\nu} \nonumber\\- \frac{e^2g^2}{2} (\beta - \alpha)^2 (\tilde{u}_\mu \tilde{\xi}^\mu)^2. \label{eq:lbroken}
\end{eqnarray}
$\tilde{\mathcal L}_{unbroken}$ describes two free, massive gauge fields. The mass of the $\tilde u$ field is given by: $m_u = g\sqrt{\langle\tilde\xi^2\rangle}$, while the mass of the $\tilde \xi$ field is $m_\xi = e\sqrt{\langle \tilde u^2\rangle}$. The equations of motion corresponding to $\tilde{\mathcal L}$ are:
\begin{eqnarray}
\partial_\mu\partial^\mu \tilde u_\nu -\partial_\nu\partial^\mu \tilde u_\mu + m_u^2 \tilde u_\nu = \frac{1}{e^2}\frac{\delta \tilde{\mathcal L}_{broken}}{\delta \tilde u^\nu},\label{eq:massivevelocity}\\
\partial_\mu\partial^\mu \tilde\xi_\nu -\partial_\nu\partial^\mu \tilde\xi_\mu + m_\xi^2 \tilde\xi_\nu = \frac{1}{g^2}\frac{\delta \tilde{\mathcal L}_{broken}}{\delta \tilde \xi^\nu}.\label{eq:massivevorticity}
\end{eqnarray}
Adopting the Lorentz gauge:
\begin{equation}
\partial_\mu \tilde u^\mu = \partial_\mu \tilde \xi^\mu = 0\,,
\end{equation}
we treat the last term in $(\ref{eq:lbroken})$ as a constraint, introducing a Lagrange multiplier to enforce it. Using the method of Lagrange multipliers, we obtain the following equivalence:
\begin{eqnarray*}
\frac{e^2g^2}{2}(\beta-\alpha)^2(\tilde u_\mu\tilde\xi^\mu)^2 + \lambda\phi^2 eg(\alpha -\beta) (\tilde u_\mu \tilde \xi^\mu) + \frac{\lambda^2\phi^4}{2} \\ 
= \frac{1}{2}\left(\lambda\phi^2 + eg(\beta-\alpha)\tilde u_\mu \tilde \xi^\mu \right)^2,
\end{eqnarray*}
where $\lambda$ describes the self-coupling of the scalar field, which acts here as a Lagrange multiplier. When $-eg(\beta-\alpha)\tilde u_\mu \tilde \xi^\mu = \lambda\langle\phi^2\rangle$, the broken Lagrangian $\mathcal L_{broken}$ becomes:
\begin{eqnarray}
\tilde L_{broken} &=& \frac{eg}{2}\gamma \tilde f_{\mu\nu}\tilde K^{\mu\nu} + e^2g(\beta^2-\alpha^2)\tilde u_\mu\tilde\xi_\nu \tilde K^{\mu\nu} \nonumber\\
&+& \lambda\langle\phi^2\rangle eg(\beta-\alpha)\tilde u_\mu\tilde\xi^\mu + \lambda^2\langle\phi^2\rangle^2.\label{eq:lunbroken}
\end{eqnarray}
Now, the two sides of $(\ref{eq:massivevorticity})$ can be separated. We set:
\begin{equation}
\partial_\mu \partial^\mu \tilde \xi_\nu + m_\xi^2 \tilde \xi_\nu = 0\,,
\end{equation}
and the right-hand side of $(\ref{eq:massivevorticity})$ becomes:
\begin{equation}
\partial_\mu\partial^\mu \tilde u_\nu + \frac{e(\beta^2-\alpha^2)}{\gamma}\tilde u_\mu\tilde  K^{\mu\nu} + \frac{\lambda\langle\phi^2\rangle(\beta -\alpha)}{\gamma}\tilde u_\nu = 0.\label{eq:reu}
\end{equation}

In the non-relativistic limit, let 
\begin{equation}
\tilde u_\mu = e^{im t}u_\mu, \quad m = \sqrt{\frac{\lambda\langle\phi^2\rangle(\beta-\alpha)}{\gamma}},\quad u_0 = 0,
\end{equation}
where $u_\mu$ evolves much more slowly than $e^{imt}$. The imaginary part of $(\ref{eq:reu})$ in the non-relativistic form is:
\begin{eqnarray}
2m\cos(mt)\partial_t u_i &-&\sin(mt)\nabla^2 u_i \nonumber\\&+&  \sin(2mt)\frac{e(\beta^2-\alpha^2)} {\gamma}u_j K^{ji}= 0.\label{eq:nonreu}
\end{eqnarray}
When:
\begin{equation}
\frac{e(\beta^2-\alpha^2)}{\gamma} = \frac{m}{\sin(mt)}, \quad \nu \equiv \frac{\tan(mt)}{2m},
\end{equation}
$(\ref{eq:nonreu})$ corresponds to the incompressible Navier-Stokes equation $(\ref{eq:vectvelocity})$ without external force or pressure. Similarly, the non-relativistic limit of $(\ref{eq:massivevelocity})$ gives the corresponding vorticity equation $(\ref{eq:vectvorticity})$.

From the above derivation, starting with an $SU(2)$ gauge theory where polarization symmetry breaks to $U(1)\times U(1)$, and taking the non-relativistic limit of the broken phase, we obtain the effective Lagrangian (\ref{eq:efflag}). In this process, we did not introduce specific dynamics for the scalar field but only considered the broken forms of the gauge fields. Similar to the BCS theory of superfluidity, the dynamics of the scalar field can characterize the system dynamics after symmetry breaking. When the scalar and vector fields decouple, the decoupling conditions may precisely correspond to the Clebsch potential flow representation.

\section{Scalar Field Theory}

In Meng and Yang's work\cite{MY2023}, quaternionic bosonic wave functions were applied to solve the Euler equations in quantum computational settings. We will demonstrate that these quaternionic wave functions are, in fact, isomorphic to vacuum fields with spontaneous symmetry breaking in gauge theory.

\subsection{Isomorphism between Quaternions and $su(2) \times u(1)$}

A wave function expressed in quaternions takes the form $\phi = a + ib + jc + kd$, with the following algebraic relations:
\begin{equation}
    ij = -ji = k, \quad jk = -kj = i, \quad ki = -ik = j \,.
\end{equation}
The generators of the $su(2) \times u(1)$ Lie algebra are:
\begin{align}
\tau^0 = \frac{\mathbb{I}_{2 \times 2}}{2}, \quad \tau^a = \frac{\sigma^a}{2}, \quad a = 1, 2, 3 \,,
\end{align}
where $\mathbb{I}_{2 \times 2}$ is the $2 \times 2$ identity matrix, and $\sigma^a$ are the Pauli matrices given by:
\begin{align}
\sigma^1 = 
\begin{pmatrix}
0 & 1 \\
1 & 0
\end{pmatrix}, \quad
\sigma^2 = 
\begin{pmatrix}
0 & -i \\
i & 0
\end{pmatrix}, \quad
\sigma^3 = 
\begin{pmatrix}
1 & 0 \\
0 & -1
\end{pmatrix}.
\end{align}
These matrices satisfy the following anticommutation and commutation relations:
\begin{align}
\{ \sigma^a, \sigma^b \} = 2 \delta^{ab} \mathbb{I}_{2 \times 2}, \quad [\sigma^a, \sigma^b] = 2i \epsilon^{abc} \sigma^c \,,
\end{align}
where $\epsilon^{abc}$ is the Levi-Civita symbol. The $su(2) \times u(1)$ Lie algebra can be expressed as:
\begin{align}
[\tau^0, \tau^0] = [\tau^0, \tau^a] = 0, \quad [\tau^a, \tau^b] = i\epsilon^{abc} \tau^c \,.
\end{align}
The quaternion representation is algebraically isomorphic to the $su(2) \times u(1)$ algebra. This is illustrated with the following examples.

The complex conjugate of the quaternion $\phi$ is given by $\bar{\phi} = a - ib - jc - kd$, which satisfies:
\begin{align}
\bar\phi \phi = a^2 + b^2 + c^2 + d^2 \,.
\end{align}
We further compute $\bar\phi i \phi$:
\begin{equation}
\bar\phi i \phi = (a^2 + b^2 - c^2 - d^2) i + 2(bc - ad) j + 2(ac + bd) k \,. \label{eq:pip}
\end{equation}

In the $su(2) \times u(1)$ basis, the wave function can be expressed as:
\begin{align}
\tilde\phi = \phi^A \tau^A = \frac{a}{2} + b \tau^1 + c \tau^2 + d \tau^3 \,.
\end{align}
It is evident that this is a Hermitian representation, as $\tilde\phi^\dagger = \tilde\phi$. Calculating the trace yields:
\begin{align}
2 \text{Tr}(\tilde\phi^{\dagger A} \tau^A \tilde\phi^B \tau^B) = a^2 + b^2 + c^2 + d^2 = \bar\phi \phi \,.
\end{align}
We also compute:
\begin{equation}
4 \tilde\phi \tau^1 \phi = (a^2 + b^2 - c^2 - d^2) \tau^1 + 2(bc - ad) \tau^2 + 2(bd + ac) \tau^3 \,.\label{eq:su2pip}
\end{equation}
It can be observed that Eq.~\eqref{eq:su2pip} matches Eq.~\eqref{eq:pip} under the quaternionic representation as in \cite{MY2023}.

\subsection{Non-Abelian Higgs Theory}

The quaternionic wave function corresponds to a scalar field with $su(2) \times u(1)$ Lie algebra indices. This algebraic isomorphism suggests the existence of a gauge theory. Consider the classical Lagrangian:
\begin{equation}
\mathcal{L} = \text{Tr} \left(\mathcal{D}_\mu \phi^* \mathcal{D}^\mu \phi - \frac{1}{4} F_{\mu\nu} F^{\mu\nu} \right) - V(\phi, \phi^*) \,, \label{eq:nonabelianHiggs}
\end{equation}
where the covariant derivative is $\mathcal{D}_\mu = \partial_\mu - g \tilde{U}_\mu^A \tau^A$, and the field strength tensor is:
\begin{align}
F_{\mu\nu} = \partial_\mu \tilde{U}_\nu^A \tau^A - \partial_\nu \tilde{U}_\mu^A \tau^A - g f^{ABC} \tau^C \tilde{U}_\mu^A \tilde{U}_\nu^B \,.
\end{align}
Expanding the kinematic term of the scalar field:
\begin{align}
\text{Tr}\left(\mathcal{D}_\mu \phi \mathcal{D}^\mu \phi^* \right) &= \partial_\mu \phi \partial^\mu \phi^* - g f^{ABC} \tilde{U}_\mu^A \phi^B \partial_\mu \phi^{*,C} \nonumber \\
&\quad + g f^{ABC} \tilde{U}_\mu^A \partial_\mu \phi^B \phi^{*,C}  \nonumber\\
&\quad + g^2 \tilde{U}_\mu^A \tilde{U}^{\mu,B} \phi^C \phi^{*,D} f^{ACE} f^{BDE} \,.
\end{align}
The Euler-Lagrange equation for the gauge field $\tilde{U}^A_\nu$ is:
\begin{align}
\partial_\mu \partial^\mu \tilde{U}^A_\nu - \partial^\mu \partial_\nu \tilde{U}_\mu^A &- g f^{ABC} \left(\phi^B \partial_\nu \phi^{*,C} - \partial_\nu \phi^B \phi^{*,C} \right) \nonumber \\
&+ 2g^2 f^{ACE} f^{BDE} \tilde{U}_\nu^B \phi^C \phi^{*,D} = 0 \,.
\end{align}
In the decoupling limit, where the coupling between the scalar and gauge fields vanishes, we find:
\begin{equation}
\tilde{U}_\nu^A = \frac{1}{2g} \frac{f^{BAC} \left(-\partial_\nu \phi^A \phi^{*,C} + \phi^A \partial_\nu \phi^{*,C} \right)}{f^{BCE} f^{ADE} \phi^C \phi^{*,D}} \,. \label{eq:decouplecond}
\end{equation}
This limit not only decouples the fields but also provides a Clebsch potential representation for fluid velocity and vorticity. The scalar field's equation of motion becomes:
\begin{equation}
\mathcal{D}_\mu \mathcal{D}^\mu \phi - \frac{\delta V(\phi, \phi^*)}{\delta \phi^*} = 0 \,.
\end{equation}

In the non-relativistic limit, where $\phi \rightarrow e^{imt} \Psi(t, \mathbf{x})$, we obtain:
\begin{eqnarray}
\left( im \partial_t - m^2 - 2g (\mathbf{u} \cdot \nabla) - \nabla^2 + g^2 \mathbf{u} \cdot \mathbf{u} \right) \Psi(t, \mathbf{x}) \nonumber\\= \frac{\delta V(\phi, \phi^*)}{\delta \phi^*} \bigg|_{\phi \rightarrow e^{imt} \Psi} \,. \label{eq:precessionNS}
\end{eqnarray}
When $\nu = \frac{1}{m}$ and $m = -2g$, Eq.~\eqref{eq:precessionNS} takes a form similar to the Schrödinger-Poisson equation, analogous to those in Gross-Pitaevskii theory, BCS theory, and the quantum computational fluid dynamics of Meng and Yang \cite{MY2023}.

\section{Conclusions and Outlook}

\subsection{Main Results and Physical Interpretations}

Through the derivations in this paper, we have obtained the following key results:

1. \textbf{The Navier-Stokes equations as a subsystem approximating the fluid system}: The NS equations describe a subsystem of a larger conservative fluid system. While energy appears dissipative within this subsystem, total energy and momentum are conserved within the overarching system.

2. \textbf{Introduction of adjoint vorticity fields and conservation mechanisms}: The adjoint vorticity field provides a mechanism for energy conservation. Through its coupling with the velocity field, it ensures the conservation of total energy in the larger system. It is also believed that the negative viscosity behavior of the adjoint field regulates vortex generation and diffusion.

3. \textbf{Derivation of light-cone vorticity equations}: The light-cone vorticities $ \boldsymbol{\eta}_\pm $ unify the vorticity and adjoint vorticity fields within a coupled framework, illustrating the complex interactions between the velocity and adjoint fields. This offers new insights into nonlinear feedback mechanisms in fluid systems.

4. \textbf{Gauge symmetry breaking mechanism}: The $SU(2)$ gauge symmetry breaking into $U(1) \times U(1)$ is identified as a polarization breaking mechanism, distinct from conventional spontaneous symmetry breaking (SSB). Further research is needed to explore this type of symmetry breaking.

5. \textbf{Complete theoretical framework}: Incorporating scalar fields, the ANOZ theory provides a comprehensive framework for fluid dynamics. This theory serves as a non-Abelian extension of the BCS theory for superfluid ${}^3\text{He}$.

\subsection{Outlook: Future Research Directions}

1. \textbf{In-depth study of polarization breaking}
The proposed \textbf{dynamical polarization breaking} mechanism differs from conventional spontaneous symmetry breaking, illustrating how gauge fields evolve to form a coupled system of dual vector fields through polarization. Future research can explore the properties of this polarization breaking both theoretically and through numerical simulations, as well as investigate its applications in other physical systems.

2. \textbf{Integrability and analytical solutions of the larger system}
Given the complex nonlinear feedback in the coupled system of velocity and adjoint vorticity fields, future work can explore the \textbf{integrability} of the larger system. By identifying conserved quantities, it may be possible to find analytical solutions, providing new theoretical insights into the formation of turbulence and vortex structures.

3. \textbf{Numerical simulations and experimental validation}
The theoretical framework presented here offers a new way to describe incompressible fluids. Future studies can validate the key conclusions through \textbf{numerical simulations} and \textbf{experimental research}. In particular, the complex interactions described by the light-cone vorticity equations provide new approaches to investigating turbulence, vortex generation, and energy transfer. Determining how to verify the dynamics of the adjoint vorticity field both numerically and experimentally presents a significant challenge. Although the adjoint vorticity field currently appears to be a theoretical construct, its physical role is analogous to that of phonons in solid-state physics. Therefore, it is highly likely to be an objectively existing physical field rather than merely a mathematical tool.

4. \textbf{Extension and applications of ANOZ theory}
As a non-Abelian extension of the BCS theory, the ANOZ theory reveals deep connections between scalar fields, broken gauge symmetries, and fluid dynamics. It offers systematic tools for the geometric formulation of fluid mechanics and serves as a complete foundational framework for describing incompressible fluids. Solving this theory lays the groundwork for a systematic approach to solving the NS equations. Understanding the integrability and solvability of the ANOZ theory is essential for advancing solutions to the NS equations. Additionally, the ANOZ theory after symmetry breaking includes two vector fields and one scalar field. The two vector fields are coupled to each other, forming the classical fluid dynamics framework of the velocity field and the associated vorticity field. The scalar field, however, is decoupled from the vector fields, representing a classical/quantum decoupling. After decoupling, the quantum nature of the scalar field remains intact. This provides a physical conjecture to the classical-quantum correspondence between quantum HSE and fluid dynamics, which established by the Clebsch mapping, or equivalently, by scalar-vector decouple condition after symmetry breaking of ANOZ theory. This classical/quantum correspondence warrants further investigation in future work.

\section*{Acknowledgements}
This work was supported by the National Natural Science Foundation of China (Grant No. 12425208).

\bibliography{references.bib}
\end{document}